\documentclass[12pt]{article}

\usepackage{graphicx,  amssymb, array, amsmath, float}

\usepackage{color}
\usepackage[authoryear]{natbib}
\usepackage{bm}
\usepackage{graphicx}
\usepackage{float}
\usepackage{subcaption}
\usepackage{multicol}
\usepackage{xr}
\usepackage{mathtools}
\usepackage{comment}
\usepackage{nameref}
\usepackage{hyperref}
\newcommand{\blind}{0}

\begin{document}

\def\spacingset#1{\renewcommand{\baselinestretch}%
{#1}\small\normalsize} \spacingset{1}

\if0\blind
{
\title{\bf A Bayesian Statistics Course for Undergraduates:\\
Bayesian Thinking, Computing, and Research}

\author{Jingchen Hu\thanks{
    The author gratefully acknowledges the Mathematics and Statistics Department at Vassar College for their support and encouragement to our experiment of Bayesian education at the undergraduate level. The author thanks the Liberal Arts Collaborative for Digital Innovation (LACOL), for their generous funding support to share this course across 10 campuses, resulting in wider student audience and richer pedagogy experience for the author. The author also thanks Jim Albert, a co-author of the recently published undergraduate-level textbook, {Probability and Bayesian Modeling} (\url{https://monika76five.github.io/ProbBayes/}, for his experience of Bayesian education and generosity to mentor junior Bayesian educators.}\hspace{.2cm}\\
    Mathematics and Statistics Department, Vassar College, \\
    Box 27, 124 Raymond Ave, Poughkeepsie, NY 12604, \\
    jihu@vassar.edu.}
\maketitle
} \fi

\if1\blind
{
  \bigskip
  \bigskip
  \bigskip
  \begin{center}
    {\LARGE\bf A Bayesian Statistics Course for Undergraduates:\\
Bayesian Thinking, Computing and Research}
\end{center}
  \medskip
} \fi

\bigskip
\begin{abstract}
We propose a semester-long Bayesian statistics course for undergraduate students with calculus and probability background. We cultivate students' Bayesian thinking with Bayesian methods applied to real data problems. We leverage modern Bayesian computing techniques not only for implementing Bayesian methods, but also to deepen students' understanding of the methods. Collaborative case studies further enrich students' learning and provide experience to solve open-ended applied problems. The course has an emphasis on undergraduate research, where accessible academic journal articles are read, discussed, and critiqued in class. With increased confidence and familiarity, students take the challenge of reading, implementing, and sometimes extending methods in journal articles for their course projects. 
\end{abstract}

\noindent%
\vfill

\newpage
\spacingset{1.45} 

\section{Introduction}
\label{intro}

Statistics educators have been actively introducing Bayesian topics into the undergraduate and graduate statistics curriculum for the past few decades. At the Joint Statistical Meetings in 1996, the Section on Statistical Education organized an invited session on the advantages, disadvantages, rationale, and methods for teaching an introductory statistics course from a Bayesian perspective. {\it The American Statistician} subsequently published a series of papers and discussion on these topics \citep{Berry1997TAS, Albert1997TAS, Moore1997TAS}. \citet{Moore1997TAS}, in particular, argued that it was premature to teach the ideas and methods of Bayesian inference in an introductory statistics course. The obstacles presented include: 1) Bayesian techniques were little used, 2) Bayesians had not yet agreed on standard approaches to standard problem settings, 3) the requirement of conditional probability can be confusing to beginners, and 4) the teaching and learning of Bayesian inference might impede the trend toward experience with real data and a better balance among data analysis, data production, and inference. 

Indeed, prior to the invention and development of the Gibbs sampler and other Markov chain Monte Carlo (MCMC) algorithms in the late 1980s and early 1990s, not only the teaching in the classroom, but also the practice of Bayesian methods, had been very limited. Nevertheless, statistics educators made great effort to innovate, especially to connect to real data problems and apply Bayesian methods to solve these problems \citep{Franck1988TAS}. 

Thanks to the revolutionary computational development and the rapid spread of Bayesian techniques used in applied problems, the teaching and learning of Bayesian methods had taken off, albeit mostly at the graduate level. Not only there are a number of Bayesian courses in statistics graduate programs, the list of available textbooks keeps growing: {\it Bayesian Data Analysis} \citep{BDA1995book}, {\it A First Course in Bayesian Statistical Methods} \citep{Hoff2009book}, {\it Bayesian Essentials with R} \citep{BayesianEss2014book}, {\it Statistical Rethinking} \citep{Rethinking2020book}, and {\it Bayesian Statistical Methods} \citep{BayesianModeling2019book}, among others. Non-statistics graduate programs, including marketing and business, cognitive science, and experimental psychology, have seen flourishing developments of Bayesian education \citep{MarketingBusiness2005book, Kruschke2014book, BCM2014book}. Some of these textbooks are suitable for undergraduate statistics students as well, for example, {\it Doing Bayesian Data Analysis} \citep{Kruschke2014book}. Statisticians have also contributed to Bayesian education for non-statisticians \citep{Gelman2008TAS, UttsJohnson2008TAS}.

How about Bayesian education at the undergraduate level? Most of the educational innovation is taking place in the introductory statistics courses, where Bayesian inference is one of the many topics. Most recently, \citet{EadieHuppenkothenSpringfordMccormick2019JSE} designed an active-learning exercise of Bayesian inference with m\&m's, and \citet{BarcenaGarinMartinTusellUnzueta2019JSE} designed a web simulator to teach Bayes theorem, with application to the search for the nuclear submarine, USS Scorpion, in 1968. Educational innovation for advanced-level undergraduate statistics courses, where Bayesian inference is typically covered as a topic in a statistical inference / mathematical statistics course, includes \citet{KuindersmaBlais2007TAS}, who designed a teaching tool of Bayesian model comparison used in a physics application of a three-sided coin, and \citet{RouderMorey2019TAS}, who proposed teaching Bayes' theorem by looking at strength of evidence as predictive accuracy. 

We are among the many Bayesian statistics educators, who believe in the huge benefits of introducing Bayesian methods into the undergraduate statistics curriculum, beyond being only a topic in introductory or statistical inference / mathematical statistics courses \citep{Witmer2017TAS, ResamplingR2018}. In this article, we propose a semester-long Bayesian statistics course for undergraduates with a background of multivariable calculus and probability. The main learning objectives are: students are expected to 1) understand basic concepts in Bayesian statistics, including Bayes' rule, prior, posterior, and posterior predictive distributions, and 2) apply Bayesian inference approaches to scientific and real-world problems. To achieve these learning objectives, we emphasize the cultivation of Bayesian thinking, the role of statistical computing and the use of real data. Furthermore, our proposed course has three important components: case studies, discussing and critiquing journal articles, and course projects. 

In Section \ref{overview}, we provide a course overview, where we introduce the pre-requisites and students' background, leading to a discussion of the choice of topics in the course. As Bayesian computing is an important and interwoven aspect of this course, Section \ref{overview:computing} describes our choices, approaches, and philosophy of computing in the course. We then proceed to the details of the proposed course in Section \ref{features}, including the three key components of the course: case studies, discussing and critiquing journal articles, and course projects, where we provide recommendations based on our experience. We also discuss assessments in the course in this section. The article ends with a discussion of students' experience, challenges, and future ideas with Section \ref{epilogue}. Supplementary materials include a course schedule, sample in-class R scripts and computing labs in R Markdown, sample homework, sample case studies, and a reading guide for a journal article. These are available in the Supplementary Materials online.

\section{Course Overview}
\label{overview}

Our proposed course is a popular elective for students pursuing statistics major or minor at Vassar College. We meet twice a week for 13 weeks, with a total of 26 class meetings. Each class meeting lasts 75 minutes, and some lectures are used as computing labs. The course pre-requisites are multivariable calculus and probability. Among the calculus and probability topics, we emphasize familiarity with transformation of random variables and joint distributions (especially joint densities of conditionally independently and identically ($i.i.d.$) distributed random variables), as these are key skills in expression of joint posterior distributions. We provide a solid review of these topics at the beginning of the course.

A typical student in this course might have prior statistics exposure, though having taken a statistics course is not required. We also do not assume prior R experience of students, though most likely they have had some exposure. To ensure students are ready for statistical computing in the course, we assign three DataCamp\footnote{For more information about DataCamp, visit \url{datacamp.com}.} courses within the first couple of weeks of the course: Introduction to R\footnote{The course link: \url{https://www.datacamp.com/courses/free-introduction-to-r}.}, Intermediate R\footnote{The course link: \url{https://www.datacamp.com/courses/intermediate-r}.}, and Introduction to the Tidyverse\footnote{The course link: \url{https://www.datacamp.com/courses/introduction-to-the-tidyverse}.}, all of which are available through DataCamp For The Classroom\footnote{For more information about the DataCamp For The Classroom, visit \url{https://www.datacamp.com/groups/education}.}. Since we do not assume prior R experience, we have made the choice to mainly use base R instead of \texttt{tidyverse} in this course, though we have seen students with \texttt{tidyverse} background opt to use \texttt{tidyverse}.

We have two main learning objectives: students are expected to 1) understand basic concepts in Bayesian statistics, including Bayes' rule, prior, posterior, and posterior predictive distributions, and 2) apply Bayesian inference approaches to scientific and real-world problems. These course objectives motivate our choice of topics and their contents:

\begin{itemize}
\item {\bf{Bayesian inferences for a proportion and for a mean}}: Covering Bayes theorem, conjugate prior, posterior distribution, and predictive distribution, with a focus of comparison of the exact solutions versus the Monte Carlo simulation solutions in one-parameter Bayesian models.
\item {\bf{Gibbs sampler and MCMC}}: Covering multi-parameter Bayesian models, why and how a Gibbs sampler works, MCMC diagnostics, coding one's own Gibbs sampler, and using Just Another Gibbs Sampler (JAGS) for MCMC estimation.
\item {\bf{Bayesian hierarchical modeling}}: Covering why a hierarchical model is preferred in certain data analysis, how to specify a multi-stage (hierarchical) prior distribution, MCMC estimation, prediction, and analyzing pooling / shrinkage effects induced in hierarchical models.
\item {\bf{Bayesian linear regression}}: Covering how to estimate a regression model, different prior choices, MCMC estimation, and predictions. 
\end{itemize}

The above are our main topics in the course, designed to provide an adequate coverage of basic Bayesian concepts, inference methods, and computing techniques within certain applied contexts. Through case studies, discussing and critiquing journal articles, and course projects, students are exposed to a much wider range of Bayesian methods in this course: some are innovative extensions of methodologies covered in class through case studies; others are much more advanced methods students encounter in their course projects. These features are designed to provide ample time and space for students to dive into research with what they have gained to achieve what they want to do. We will provide details and discuss our choices of these features in Section \ref{features}.

As Bayesian computing is an important and interwoven aspect of any modern Bayesian course, we proceed to describe our choices, approaches, and philosophy of computing in our proposed course in Section \ref{overview:computing}.

\subsection{Computing in The Course}
\label{overview:computing}

\subsubsection{Goals and Approaches}
\label{overview:computing:goals}

At the undergraduate-level, we expect students to use computing techniques for implementing Bayesian methods in applied problems. Moreover, going through the computing aspect of Bayesian inference enhances students' understanding of the methods themselves. We achieve these two computing goals by a two-stage process:

\begin{itemize}
\item Stage 1 (first 1/3 of the course): In conjugate cases, e.g. beta-binomial, normal-normal, gamma-normal, gamma-Poisson, implement and compare exact solutions and Monte Carlo approximation solutions to posterior and predictive inference. 

\item Stage 2 (second 2/3 of the course): Introduce JAGS for implementing Gibbs samplers, compared to self-coded Gibbs samplers for simple cases. From then on, use JAGS for subsequent topics, for example, Bayesian hierarchical modeling and Bayesian linear regression (not only Gibbs samplers, but also Metropolis-Hastings algorithms).
\end{itemize}

\subsubsection{Exact Solutions versus Simulation Solutions}
\label{overview:computing:exactsim}

The focus of Stage 1 is to familiarize students with R programming and Monte Carlo techniques in simulating posterior and predictive distributions. In conjugate cases, analytical posterior and predictive distributions are available, which are great examples for students to compare the exact solutions and Monte Carlo simulation solutions. Furthermore, simulating the predictive distributions and performing posterior predictive checks give students ample opportunities to distill the essence of Bayesian computing. It is therefore desirable not to introduce JAGS or any other MCMC estimation software at this stage, avoiding the tendency to use these software as a ``black box".

\subsubsection{Why and How to Use JAGS}
\label{overview:computing:JAGS}

The shift from self-coding to available software such as JAGS takes place in covering the Gibbs sampler and MCMC diagnostics. Simple cases, such as a two-parameter normal model, are used to introduce the definition and derivation of full conditional posterior distributions, the keys to writing one's own Gibbs sampler. Given example R scripts of a Gibbs sampler (involving functions and loops), students practice writing Gibbs samplers to explore important aspects of MCMC, leading to discussions of MCMC diagnostics.

\begin{figure}[H]
\centering
\includegraphics[width=0.35\textwidth]{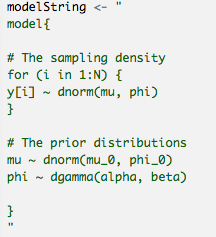}
\caption{The JAGS script to express the sampling density and the prior distributions.}
\label{fig:JAGS}
\end{figure}

Once students have a solid understanding of the mechanics of Gibbs samplers and have gained the ability to write their own Gibbs samplers, JAGS software is introduced, focusing on its descriptive nature of the specified Bayesian models and its comparison to a self-coded Gibbs sampler. To show JAGS's descriptive nature, Figure \ref{fig:JAGS} presents the JAGS script, with the expressions of the sampling density and the prior distributions to its right. To compare JAGS output to the output of a self-coded Gibbs sampler, students are prompted to revisit previously covered aspects of MCMC, further distilling the keys to MCMC and its diagnostics.

\begin{multicols}{2}

\vspace{3mm}
\begin{itemize}
\item[-] The sampling density:
\begin{equation*}
Y_1, \cdots, Y_n \mid \mu, \sigma \overset{i.i.d.}{\sim} \textrm{Normal}(\mu, \sigma).
\end{equation*}

\item[-] The prior distributions:
\begin{eqnarray*}
\mu &\sim& \textrm{Normal}(\mu_0, \sigma_0), \\
1/\sigma^2 = \phi &\sim& \textrm{Gamma}(\alpha, \beta).
\end{eqnarray*}
\end{itemize}

\end{multicols}

Recall our two aforementioned computing goals in this course: we expect students to use computing techniques for implementing Bayesian methods in applied problems, and students' understanding of the Bayesian methods should be further enhanced by going through the computing aspect of these methods. Using JAGS achieves both goals: JAGS ``frees up" students to explore non-conjugate priors and advanced Bayesian models, which involves additional MCMC samplers, such as the Metropolis-Hastings algorithm. Moreover, it enhances students' understanding of the Bayesian models being implemented: although students might not know the actual algorithms that JAGS performs, they need to be absolutely clear about how to write the JAGS script to implement the Bayesian models they have intended to use. We believe at the undergraduate-level, JAGS is sufficient and self-directed, that there is no need to provide an R package for the proposed Bayesian statistics course.

In addition to JAGS, Bayesian inference Using Gibbs Sampling (BUGS)\footnote{For more information, visit https://www.mrc-bsu.cam.ac.uk/software/bugs/} is another popular MCMC estimation software, whose syntax and usages are similar to those of JAGS. Stan\footnote{For more information, visit: https://mc-stan.org/} is rapidly growing in popularity, and many available packages provide wrapper functions such as the \texttt{stan\_glm()} in the \texttt{rstanarm} package, that shares similar syntax and usages with the standard \texttt{glm()} R function. We have made the deliberate decision to use JAGS due to its aforementioned descriptive nature of the sampling density and prior distributions, its compatibility with various operating systems, and its emphasis on understanding the Bayesian model specification (unlike the ``black box" style of Stan-based wrapper functions). We also note that, to the best of our knowledge, almost all undergraduate-level Bayesian textbooks use JAGS \citep{Kruschke2014book, BayesianModeling2019book, ProbBayes2019book}. \citet{AlbertHu2020JSE} provides a review of Bayesian computing in the undergraduate statistics curriculum and their recommendations.

%

\section{Course Features and Recommendations}
\label{features}

This section provides further details of the course. We first describe homework, computing labs, and exams in the course. To achieve our learning objectives, we incorporate three important components into our course: case studies, discussing and critiquing journal articles, and course projects. Sections \ref{features:casestudies} through \ref{features:project} present and discuss these three key components one-by-one and our recommendations for enhancing students' learning experience. In Section \ref{features:assessment}, we present all course assessment components and their percentages of the final grade, with a discussion about how they are used to assess learning objectives and assign grades.

Homework is the main assessment tool in the first 3/5 of the semester, assigned once every two weeks and 4 in total: each homework has a written portion, focused on exercises on topics such as deriving the posterior distributions, and an R portion, focused on implementing Bayesian inference methods with real data. The written portion enhances students' understanding of key Bayesian concepts, while the R portion gives students ample opportunity to practice using R for Bayesian inference. In-class computing labs are designed to enhance students' programming skills, assigned throughout the semester and 5 in total. The topic of each computing lab is closely connected to the lecture material - we will describe a computing lab designed for deepening students' understanding of a journal article in Section \ref{features:papers}. The labs are prepared and expected to be finished using R Markdown, which greatly help students gain familiarity and skills with this important tool. Two midterm exams are given. Both have an in-class portion focused on Bayesian thinking and theoretical derivations, and a take-home portion focused on Bayesian computing to implement appropriate methods to solve applied problems. A course schedule with assessment components, including homework, labs, and exams, is available in the Supplementary Materials.

We now proceed to describe the three unique and important components of our course.

\subsection{Case Studies}
\label{features:casestudies}

Like \citet{AllenbyRossi2008TAS}, we believe case studies are effective ways to illustrate applied Bayesian analysis: one constructs appropriate prior distributions, develops the likelihood, computes the posterior distributions, and finally communicates their results to address the questions of interest. In the last 2/5 of the semester, we introduce 3 case studies in place of homework assignments, where students are encouraged to explore extension of learned methods and / or create new approaches, to solve open-ended applied problems. Students are paired up and given one week to work on each case study. Some case studies are worth multiple rounds of attempts and discussions, while others are more straightforward.

The open-ended nature of case studies encourages students to build upon what they have learned and think outside of the box. Furthermore, they provide opportunities to introduce advanced modeling techniques. For example, given a dataset consisting of two clusters of students' multiple-choice test scores, are there a knowledgable group and a random guessing group? If so, can we differentiate them? Furthermore, can we make inference about some group parameters, such as the accuracy of answering a question? Given their experience, students will usually take the hierarchical modeling approach with a pre-determined set of two groups. After a first round of case study reports and discussion, we point out a clear drawback of the approach: are the pre-determined groups are reasonable, especially for those borderline scores (e.g. accuracy rates of 60\% and 70\%, which are higher than the random guessing of 50\% but lower than the knowledgable 90\%)? Latent class models, where no pre-determined group for each observation is assigned, are more suitable modeling techniques for such context. With an appropriate level of model introduction and sample R / JAGS syntax, students re-take this case study and learn by themselves the details of the model and estimation process. Case studies like this one could introduce important applied problem and new Bayesian methods simultaneously, further strengthening students' learning. Moreover, they could showcase and create discussions of limitations of familiar approaches, and open up opportunities for more suitable, although inevitably more advanced modeling techniques. This case study is available in the Supplementary Materials. 

In addition to designing suitable case studies, we recommend the following practices to enrich students' learning experience. First, assign students to case studies in pairs to encourage collaborative work. If possible, allow students to work with a different partner in every case study. Second, ask pairs to upload their case study write-ups on the learning management system (LMS, such as Moodle, Canvas etc.) before class discussion. Third, during lecture, allow students to first discuss their approaches and findings in a small group, and then to discuss and critique different approaches as an entire class.

\subsection{Journal Articles}
\label{features:papers}

\citet{Cobb2015TAS}'s five imperatives to ``rethink our undergraduate curriculum from the ground up" highly resonate with us, especially the last and the most important imperative: ``teach through research". We believe undergraduate students can and should be reading academic journal articles as part of their education, provided that the articles have the right content at the right level. 

Journals such as {\it The American Statistician} are great sources of high-quality and accessible journal articles for undergraduates. For our Bayesian course, \citet{ExpGibbs1992TAS} has been our favorite. Concise and nicely written, the authors gives a simple explanation of how and why the Gibbs sampler works, illustrates its properties with a two-by-two simple case, designs simulation studies, and analyzes the results. Furthermore, as it was an early paper, some aspects of the Gibbs sampler, such as how to obtain independent parameter draws, could be different from current practice. All these provide wonderful opportunities for students to read and learn about the development of the Gibbs sampler, and to discuss and critique different practices. 

In addition to selecting articles with the right content and at the right level, we recommend the following practices to enhance students' learning experience. First, provide a reading guide containing several questions to help students navigate the article. A selection of questions of varied types - some about verifying mathematical expressions and others about describing methodology in one's own words - is a great blend. Second, ask students to post their responses to these questions on the LMS before class discussion to facilitate small-group and entire-class discussions in class. If possible, ask additional posts after class discussion. Third, if the article contains simulation studies, design a computing lab to allow students to replicate simulation studies and graphical results presented in the article. In our case with \citet{ExpGibbs1992TAS}, replicating one of their simulation studies requires students to be very clear about how many Gibbs samplers are run and how many iterations each run needs to take, which undoubtedly deepen their understanding of Gibbs samplers and enhance their computing skills. Our reading guide and computing lab are available in the Supplementary Materials. 

The benefits of reading, discussing, critiquing, and replicating journal articles go beyond the articles themselves. Through this process, students have gained confidence and skills, encouraging them to take the challenge to read, understand, implement, and sometimes extend methods from journal articles in their course projects, where we have frequently witnessed students' growth from a focus on completing assignments to a focus on conducting research.

\subsection{Course Projects}
\label{features:project}

In addition to the exposure to journal articles, our next step of implementing \citet{Cobb2015TAS}'s ``teach through research" imperative is a course project.

Students in our course are encouraged to brainstorm project ideas from day one. In addition to several instructor-selected examples of Bayesian methods solving interesting applied problems, our first lecture includes a few video clips of students' projects from previous semesters (a 2-minute introduction video of the course project is required as part of students' project submission). This immediately shifts the focus of interesting and exciting projects considered by us, the instructors, to those by them, the fellow students. The variety of project topics and interests not only showcases what students could achieve in their projects, but also motivates them to choose what they want to explore. 

Depending on students' academic background and career aspiration, there could exist clusters of interests and topics. In the case of Vassar College: we have a number of double majors of Mathematics / Statistics and Economics, leading to groups of students working on projects related to economics and finance; we have a Bayesian cognitive scientist faculty member in the Cognitive Science Program, leading to groups of students analyzing experimental data to explore learning theories. Hot topics, such as neural networks and natural language processing, inevitably attract students' attention and are reflected in their project interests and topic choices.


Here we share additional information about practices we use to form project teams, keep every project on track, and present project outcomes. Our semester is 13 weeks long. Within the first week of the semester, students are encouraged to indicate their project interests through a self introduction post on the LMS. A list of project interests is extracted from their posts and shared with all students through an editable Google Doc, which students can freely browse and add more thoughts. It is at this stage that students start to find shared interests with each other, and slowly project teams begin to form. By Week 6, students settle down on their project topics, and submit a one-page project proposal. Each project team (up to 3 students) needs to meet with the instructor before submitting the project proposal, and detailed feedback of feasibility and advice is given by Week 7, midway of a 13-week semester. From Week 8, each project team creates a weekly schedule to complete the project. There are 2 credit-bearing check points for every team: a methodology draft by Week 10 and a project draft by Week 12. 

On the last day of class in Week 13, teams present their projects at a poster session. The poster session lasts for 75 minutes. Typically we break all teams into 3 sessions, and each session is allocated with 15 minutes, with a 5-minute discussion break between sessions, and concluding with a 20-minute final discussion. This arrangement allows students to present their own posters and to explore other students' work. The 5-minute discussion break invites students to share their thoughts after learning about other students' projects. In addition, each team submits a 2-minute introduction video about their project, and every student is asked to watch the videos before the poster session. These 2-minute videos help presenters give a high-level pitch about their work. It also helps everyone to plan their poster session better, for example, to spend more time on a poster which they are curious about based on the introduction video. 

Course projects naturally grow into independent studies in the following semesters. Past and current independent study topics stemming from this Bayesian statistics course include: Bayesian estimation of future realized volatility, Bayesian inference with Python, Bayesian nonparametric models, Bayesian time series, and Bayesian variable and model selection. Students in these independent studies are engaged in almost the entire process of applied statistics research: literature review, collect / find datasets, implement methods, analyze the results, and write a journal-style article / report. Some projects can be turned into a new topic in the future iterations of this Bayesian statistics course. Moreover, working with students on topics of their interests exposes the instructor to new research areas and ideas. 

\subsection{Assessment}
\label{features:assessment}

We present Table \ref{tab:assess} with all course assessment components and their percentages of final grade. Details about when each assessment is assigned and how long students are expected to complete can be found in the course schedule, available in the Supplementary Materials.

While all other components are self-explanatory to some extent, we would like to discuss our rationale of creating the participation component. It includes DataCamp modules, case studies, and paper discussions (online and in-class), with the latter two corresponding to the two unique and important features of our course, described in Sections \ref{features:casestudies} and \ref{features:papers} respectively. They are all assessment tools of an open-ended nature. As mentioned previously, homework is assigned for the first 3/5 of the semester. During this period, students are gradually absorbing new knowledge and techniques, and homework consisting of a set of ``close-ended" questions is a great tool to assess students' learning. In the last 2/5 of the semester, students would have acquired basic knowledge and techniques. Moreover, they have been exposed to a collection of Bayesian modeling strategies, which has prepared them to solve more open-ended questions. We therefore replace traditional ``close-ended" homework with open-ended case studies, and we have deliberately made the choice of making case studies as part of participation instead of being formally graded. 

\begin{table}
\caption{Assessment components and their percentages of final grade. \label{tab:assess}}
\centering
\begin{tabular}{p{5in} p{1.2in} }
\hline
Assessment Component &  Percentage \\ \hline
Homework and labs & 25\% \\
Participation (DataCamp modules, case studies, and paper discussions) & 10\% \\
Midterm exams & 40\% (20\%$\times$2) \\
Project & 25\% \\ \hline
\end{tabular}
\end{table}

In sum, assessments of ``close-ended" nature, including homework, labs, and midterm exams are designed based on the learning objective of understanding basic concepts in Bayesian statistics, while assessments of open-ended nature, including case studies, paper discussions, and projects are designed based on the learning objective of applying Bayesian inference approaches to scientific and real-world problems.

\section{Epilogue and Discussion}
\label{epilogue}

The various learning activities, tasks, and assessments in our proposed course fit well with many of the recommendations of the GAISE Report \citep{GAISE2016}: integrate real data with a context and a purpose (case studies and projects), foster active learning (computing labs, journal articles, case studies, and projects), use technology to explore concepts and analyze data (R and JAGS), and use assessments to improve and evaluate student learning (a well-designed hybrid of ``close-ended" and open-ended assessment tools).

Students' experience - through informal and formal evaluations - has been overall positive. Despite a challenging and heavy workload, students recognize their knowledge building and skills building in this course. Many have expressed positive experience with the course project, and the structured weekly schedule has been highly appreciated. 

Since the introduction of the course in Fall 2016, we have been running it once every academic year at Vassar College. Our emphasis on student research has been proved successful: so far we have had two 1st place winners in the intermediate statistics category of the Undergraduate Class Project Competition (USCLAP), organized by the Consortium for the Advancement of Undergraduate Statistics Education (CAUSE) and the American Statistical Association (ASA)\footnote{For more information about the USCLAP, visit \url{https://www.causeweb.org/usproc/usclap}.}. 

We recognize the challenges of teaching an undergraduate-level Bayesian statistics course for students with limited background, and we see room for improvement. For one, our course contains nontrivial R programming and statistical methods, which can be especially challenging to students with little or no R programming backgrounds. We have been supplementing with in-class R examples, computing labs, and DataCamp courses, and we see room for improvement on this end. With no requirement of linear algebra, we have little means (and little time) to cover important topics such as Bayesian model selection and variable selection in depth. We see room for improvement, and we believe it is highly likely for instructors at other institutions to have linear algebra as one of the pre-requisites, such that topics of model selection and variable selection can be better incorporated into their courses. 

The same goes for prior exposure to statistics, especially the classical / Frequentist paradigm. While Vassar College students might come into our course with no statistics background, we can easily envision instructors at other institutions to include comparisons and discussions of classical / Frequentist and Bayesian methods. These comparisons and discussions can be added into topics such as two-sample comparisons (proportion and mean), hierarchical / multi-level modeling, and linear regression.

We see the possibility and promise of applying some of our proposed pedagogy approaches to many advanced undergraduate-level statistics courses. We believe statistical computing, not only as a means to implement statistical methods, but also to strengthen students' understanding of the statistical methods, is a crucial and interwoven component of any modern statistics course. Exposing students to accessible academic journal articles enriches students' learning experience, and boosts students' confidence and critical thinking skills. Furthermore, we support ``teach through research" by course projects, and we believe in instructors' mentoring of students' course projects.

For prospective instructors, we believe the experience of a graduate-level Bayesian statistics course is sufficient, though Bayesian research experience is desirable to teach an undergraduate-level Bayesian course. We have teaching and learning materials publicly available at \url{https://github.com/monika76five/Undergrad-Bayesian-Course}. We have a recently published textbook for undergraduate Bayesian education in the CRC Texts in Statistical Science series. Details are available at \url{https://monika76five.github.io/ProbBayes/}. Interested readers can also refer to \citet{Hu2019CHANCE} for sharing this course through a hybrid model across several liberal arts colleges.

\section*{Supplementary Materials}

Please see our Supplementary Materials online for a course schedule, sample in-class R scripts and computing labs in R Markdown, sample homework, sample case studies and a reading guide for a journal article.

\section*{Acknowledgements}

We are very grateful to the editor, the associate editor, and two reviewers for their useful comments and suggestions, which we believe have improved our work. 
\bibliography{BayesEdbib}

\end{document}



\begin{center}
{\large{\bf{Supplementary Materials for \\
            A Bayesian Statistics Course for Undergraduates: \\
            Bayesian Thinking, Computing, and Research}}}\\
\vspace{5mm}
\end{center}


\tableofcontents

\clearpage
\newpage

\invisiblesubsection{{\bf{Course schedule and assessment components}}}
\summary{The final course schedule from Fall 2019.}
\includepdf[pages=-,pagecommand={}]{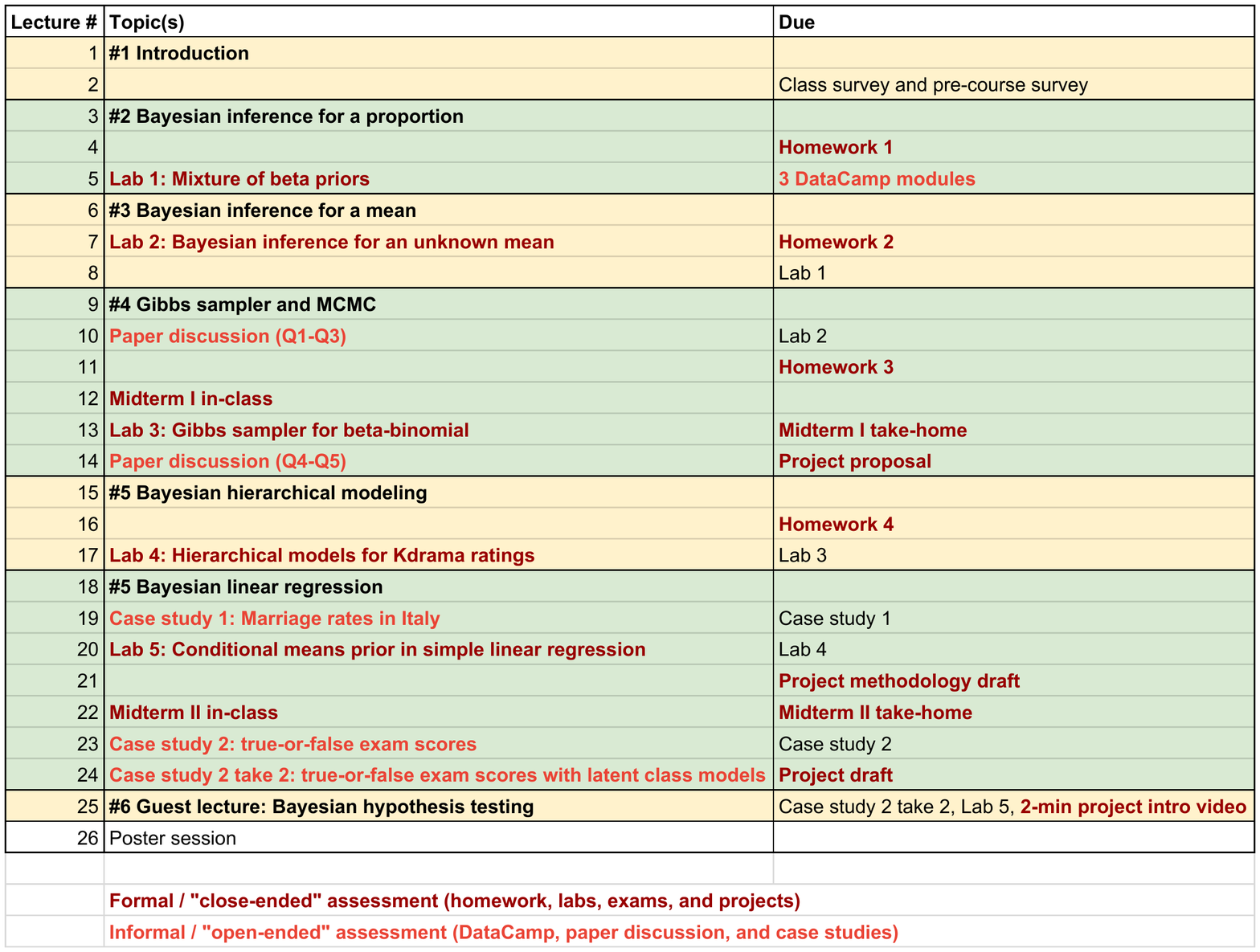}

\invisiblesubsection{{\bf{Bayesian hierarchical modeling (R scripts)}}}
\summary{Sample in-class R scripts; R Markdown output.}
\includepdf[pages=-,pagecommand={}]{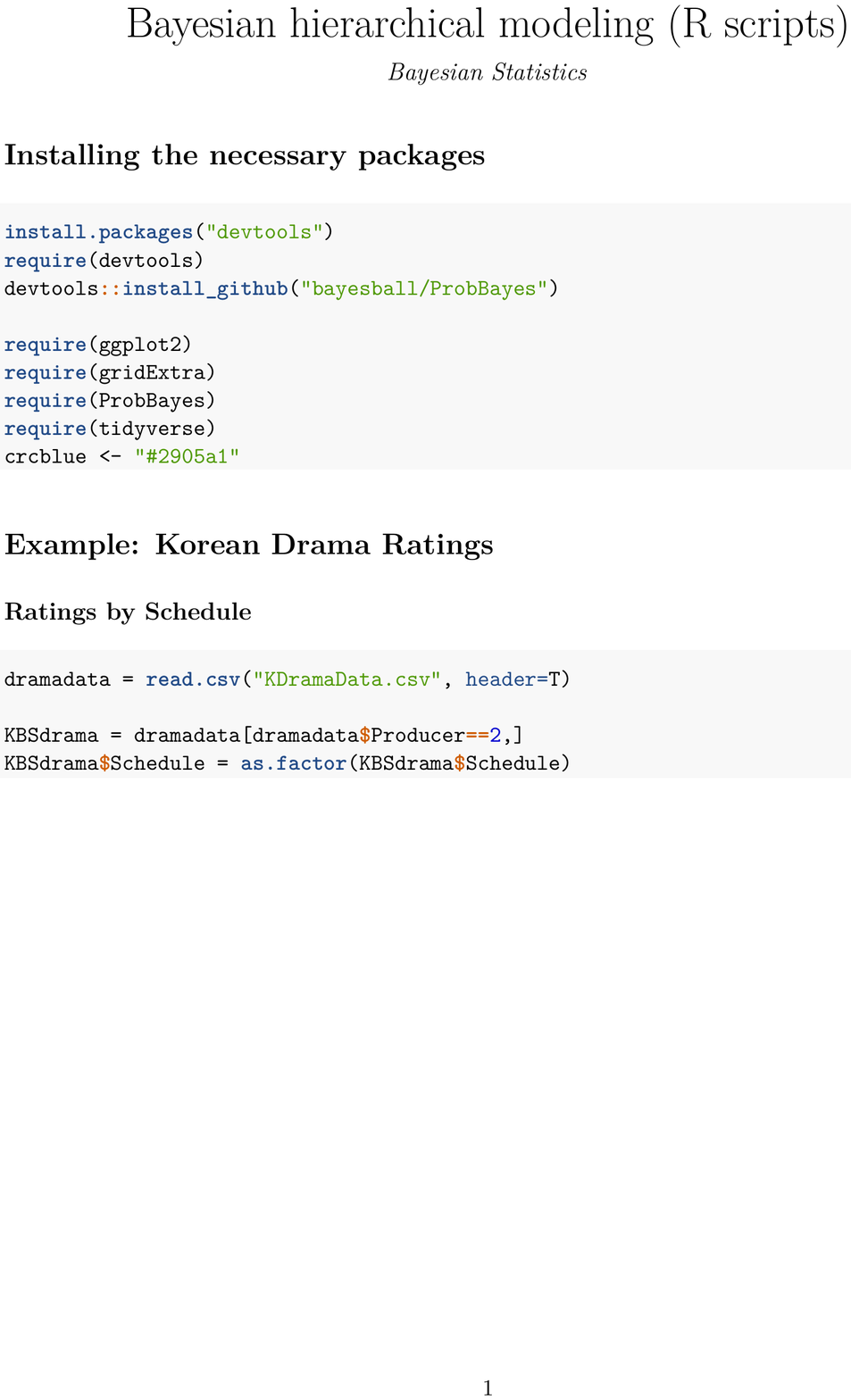}

\invisiblesubsection{{\bf{Bayesian Statistics Homework 2}}}
\summary{Sample homework assignment.}
\includepdf[pages=-,pagecommand={}]{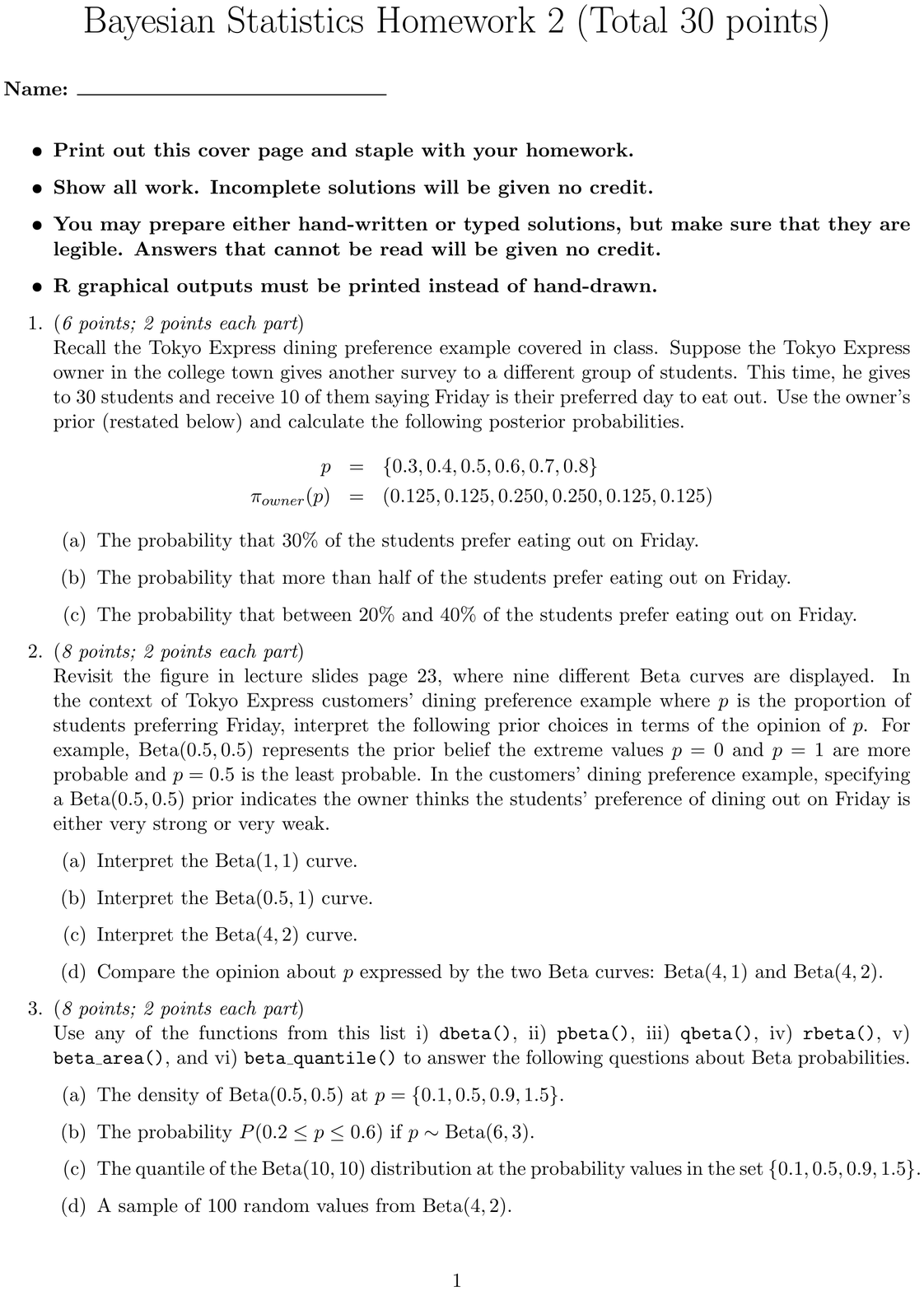}

\invisiblesubsection{{\bf{Bayesian Statistics Case Study 2}}}
\summary{Sample case study.}
\includepdf[pages=-,pagecommand={}]{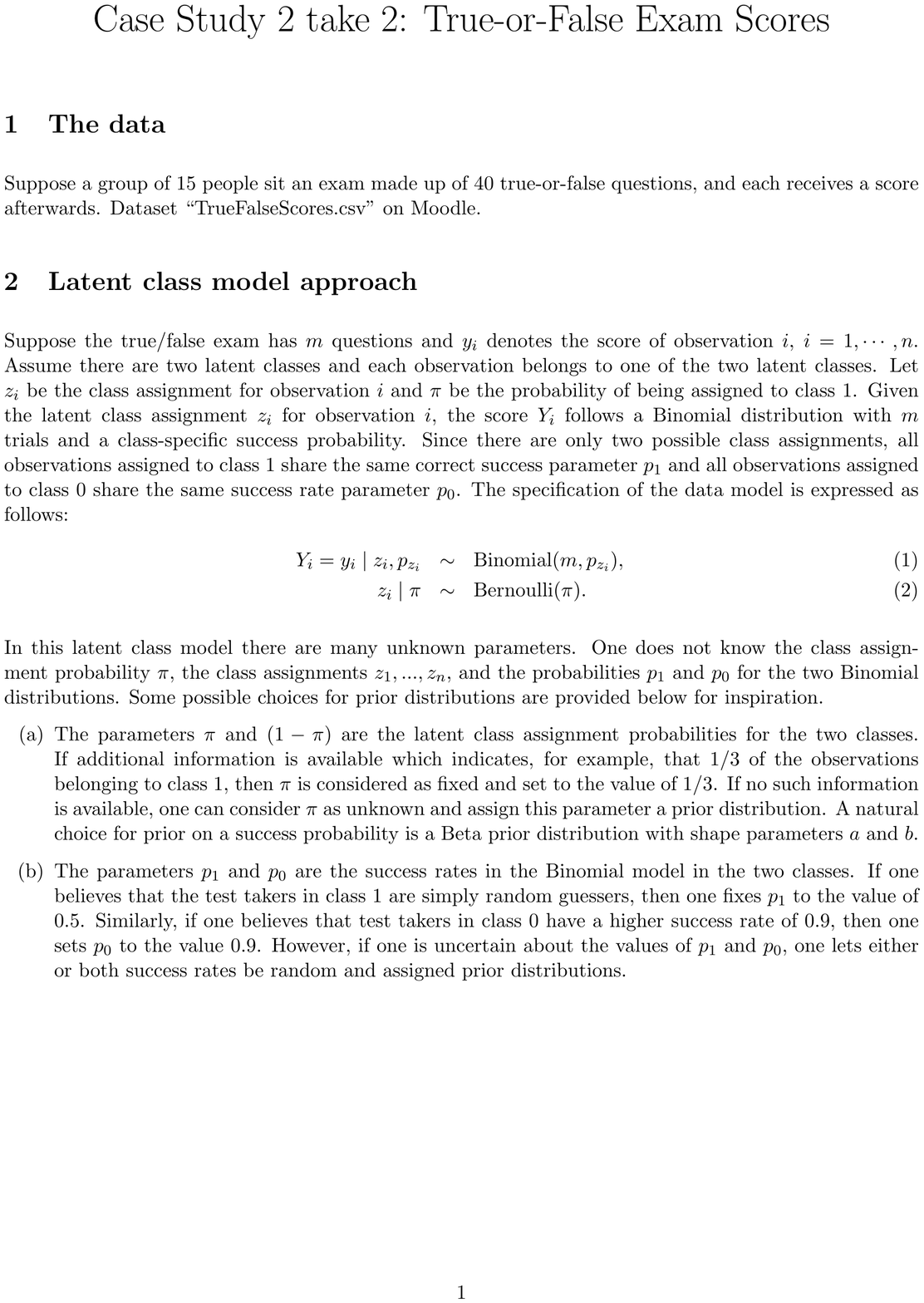}

\invisiblesubsection{{\bf{Reading Guide for Explaining the Gibbs Sampler}}}
\summary{Reading guide for Casella and George (1992).}
\includepdf[pages=-,pagecommand={}]{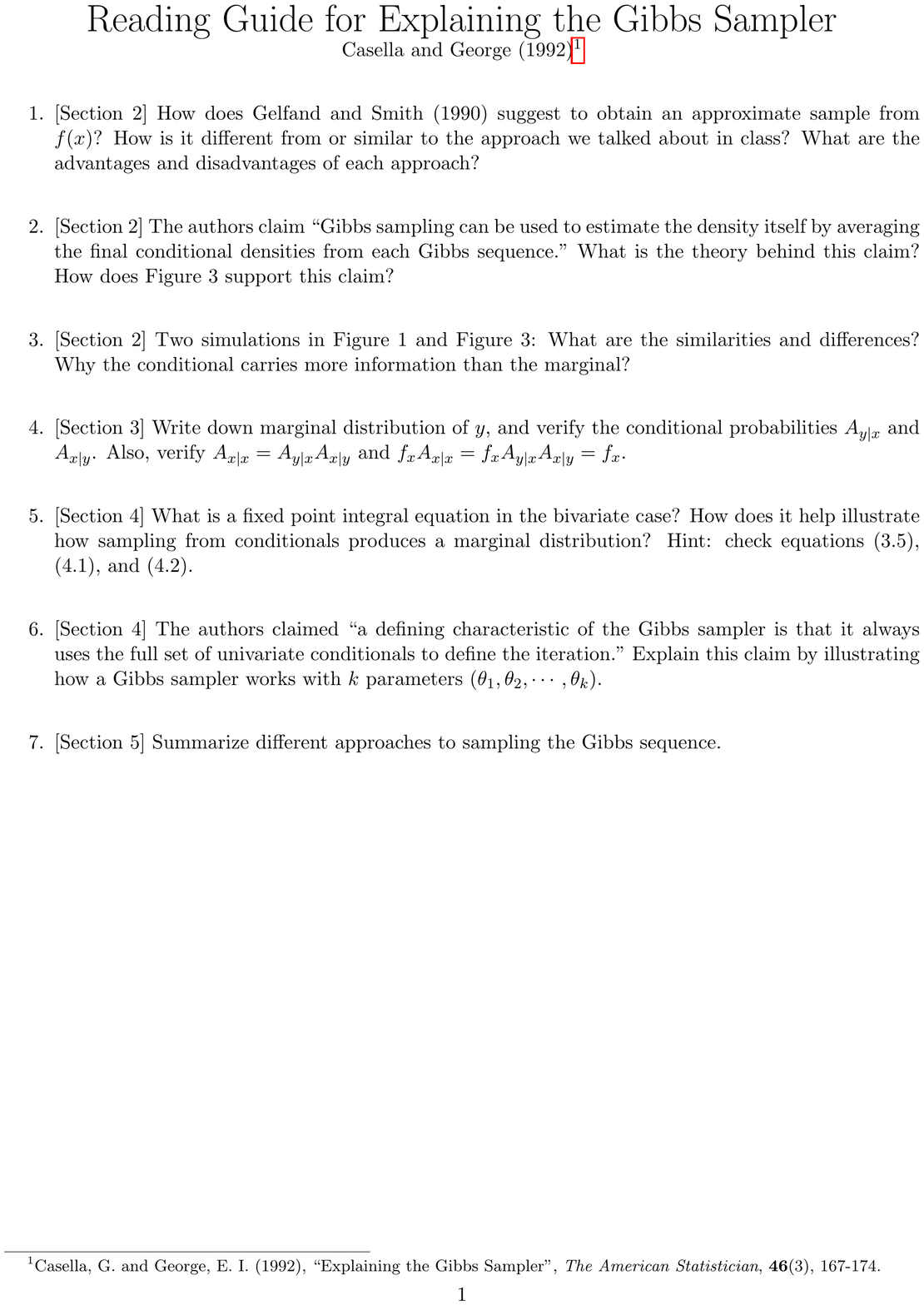}

\invisiblesubsection{{\bf{Lab 3: Gibbs Sampler of a Beta-Binomial}}}
\summary{The lab designed for replicating Casella and George (1992).}
\includepdf[pages=-,pagecommand={}]{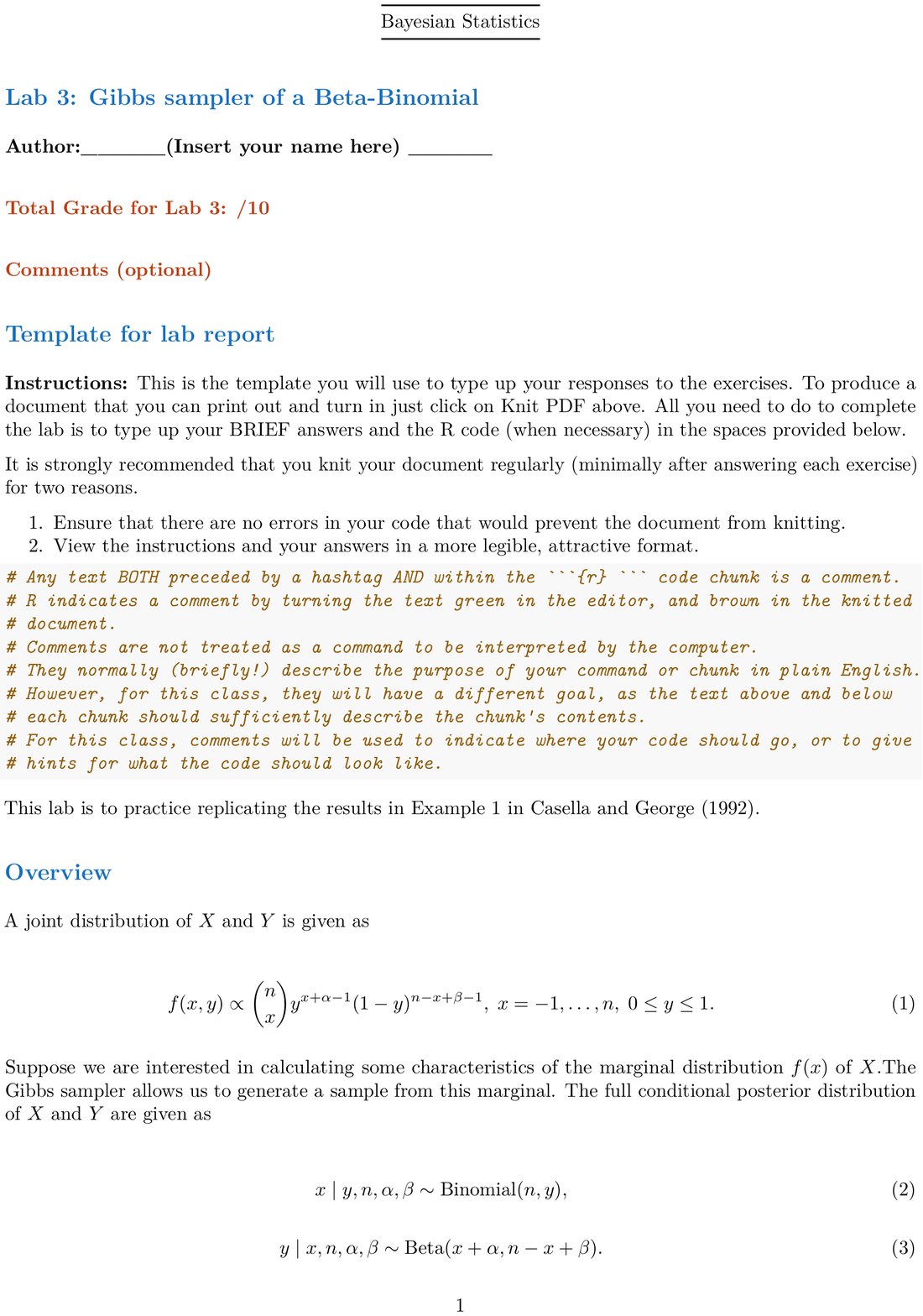}

\bibliography{BayesEdbib}